\documentstyle[preprint,aps]{revtex}
\begin{document}
\draft         
\preprint{IFUSP-P 1065,~IFT-P.053/93~and~hep-ph/9308353}
\title{Triple Vector Boson Processes in $\gamma\gamma$ Colliders}
\author{F.\ T.\ Brandt $^a$,  O.\ J.\ P.\ \'Eboli $^a$,
E.\ M.\ Gregores $^b$,
\\
M.\ B.\ Magro $^a$, P.\ G.\ Mercadante $^a$,
and S.\ F.\ Novaes $^b$}
\address{$^a$ Instituto de F\'{\i}sica,
Universidade de S\~ao Paulo, \\
C.P. 20516,  01498-970 S\~ao Paulo, Brazil \\
$^b$ Instituto de F\'{\i}sica Te\'orica,
Universidade  Estadual Paulista, \\
Rua Pamplona 145,  01405-900 S\~ao Paulo, Brazil.}
\date{\today}
\maketitle
\begin{abstract}
We study the production of three gauge bosons at the next generation
of linear $e^+e^-$ colliders operating in the $\gamma\gamma$ mode. The
processes $\gamma\gamma \rightarrow W^+W^-V$ ($V=Z^0$, or $\gamma$)
can provide direct information about the quartic gauge-boson
couplings.  We analyze the total cross section as well as several
dynamical distributions of the final state particles including the
effect of kinematical cuts.  We find out that a linear $e^+e^-$
machine operating in the $\gamma\gamma$ mode will produce 5--10 times
more three-gauge-boson states compared to the standard $e^+e^-$ mode
at high energies.
\end{abstract}


\newpage


\section{Introduction}
\label{sec:int}

The multiple vector-boson production will be a crucial test of the gauge
structure of the Standard Model since the triple and quartic vector-boson
couplings involved in this kind of reaction are strictly  constrained by
the $SU(2)_L \otimes U(1)_Y$ gauge invariance.  Any small deviation from
the Standard Model predictions for these couplings spoils the intimate
cancellations  of the high energy behaviour between the various
diagrams, giving  rise to an anomalous growth of the cross section with
energy.  It is important to measure the vector-boson selfcouplings
and look for deviations from the Standard Model, which would provide
indications for a new physics.

The production of several vector bosons is the ideal place to search
directly for any anomalous behaviour of the triple and quartic
couplings.  The reaction $e^+ e^- \rightarrow W^+ W^-$ will be
accessible at LEP200 and important information about the $WW\gamma$
and $WWZ$ vertices will be available in the near future
\cite{ano:ee}. Nevertheless, due to its limited center of mass energy
available, we will have to wait for colliders with higher center of mass
energy in order to produce a final state with three or more gauge bosons
and to test the quartic gauge-boson coupling. The measurement of the
three-vector-boson production cross section can provide a non-trivial
test of the Standard Model that is complementary to the analyses of the
production of vector-boson pairs.  Previously, the cross sections for
triple gauge boson production in the framework of the Standard Model
were presented for $e^+e^-$ colliders \cite{bar:plb,bar:num,gunion} and
hadronic colliders \cite{bar:plb,golden}.

An interesting option that is deserving a lot of attention nowadays is
the possibility of transforming a linear $e^+e^-$ collider in a
$\gamma\gamma$ collider. By using the old idea of  Compton laser
backscattering \cite{las0}, it is possible to obtain very energetic
photons from an electron or positron beam. The scattering of a laser
with few GeV against a electron beam is able to give rise to a scattered
photon beam carrying almost all the parent electron energy with similar
luminosity of the electron beam \cite{laser}. This mechanism can be
employed in the next generation of $e^+e^-$ linear colliders
\cite{pal,bur} (NLC) which will reach a center of mass energy of 500--2000 GeV
with a luminosity of $\sim 10^{33}$ cm$^{-2}$ s$^{-1}$. Such machines
operating in $\gamma\gamma$ mode will be able to study multiple vector
boson production with high statistic.

In this work, we examine the production of three vector
bosons in $\gamma\gamma$ collisions  through the reactions
\begin{equation}
\eqnum{I}
\label{z}
\gamma + \gamma \rightarrow W^+  +  W^-  + Z^0 \; ,
\end{equation}
\begin{equation}
\eqnum{II}
\label{g}
\gamma + \gamma \rightarrow W^+  +  W^- + \gamma \; .
\end{equation}
These processes involve only interactions of between the gauge bosons
making more evident any deviation from predictions of the Standard Model
gauge structure. Besides that, there is no tree-level contribution
involving the Higgs boson which eludes all the uncertainties coming from
the scalar sector, like the Higgs boson mass. Nevertheless, the
production of multiple longitudinal gauge bosons can shed light on the
symmetry breaking mechanism even when there is no contribution coming
from the standard Higgs boson. For instance, in models where the
electroweak-symmetry breaking sector is strongly interacting there is an
enhancement of this production \cite{golden,strong}.

We analyze the total cross section of the processes above, as well as
the dynamical distributions of the final state vector bosons. We
concentrate on final states where the $W$ and $Z^0$ decay into
identifiable final states. We conclude that for a center of mass
energy $\sqrt{s} \gtrsim 500$ GeV and an annual integrated luminosity
of 10 fb$^{-1}$, there will be a promising number of fully
reconstructible events. Moreover, we find out that a linear $e^+e^-$
machine operating in the $\gamma\gamma$ mode will produce 5--10 times
more three-gauge-boson states compared to the standard $e^+e^-$ mode
at high energies.

The outline is as follows. In  Sec.\ \ref{sec:res}, we introduce the
laser backscattering spectrum,  and present the details of the
calculational method. Section \ref{cs:dis} contains our results for the
total cross section and the kinematical distributions of the final state
gauge bosons for center of mass energies $\sqrt{s} = 0.5$ and $1$ TeV.
This paper is supplemented by an appendix which gives  the invariant
amplitudes for the above processes.

\section{Calculational Method}
\label{sec:res}

The cross section for the triple-vector-boson production via
$\gamma\gamma$ fusion can be  obtained by folding the
elementary cross section  for the subprocesses $\gamma\gamma
\rightarrow WWV$ ($V= Z^0,~ \gamma$) with the photon
luminosity ($dL_{\gamma\gamma}/dz$),
\begin{equation}
d\sigma (e^+e^-\rightarrow \gamma\gamma \rightarrow WWV)(s) =
\int_{z_{\text{min}}}^{z_{\text{max}}} dz ~ \frac{dL_{\gamma\gamma}}{dz} ~
d \hat\sigma  (\gamma\gamma \rightarrow WWV) (\hat s=z^2 s) \; ,
\end{equation}
where $\sqrt{s}$ ($\sqrt{\hat{s}}$) is the $e^+e^-$
($\gamma\gamma$) center of mass energy and $z^2= \tau \equiv
\hat{s}/s$. Assuming that the whole electron beam is converted
into photons via the laser backscattering mechanism, the relation
connecting the photon structure function $F_{\gamma/e} (x,\xi)$ to the
photon luminosity is
\begin{equation}
\frac{d L_{\gamma\gamma}}{dz} = 2 ~ \sqrt{\tau}  ~
\int_{\tau/x_{\text{max}}}^{x_{\text{max}}} \frac{dx}{x}
F_{\gamma/e} (x,\xi)F_{\gamma/e} (\tau/x,\xi) \; .
\label{lum}
\end{equation}
For unpolarized beams the photon-distribution function \cite{laser} is
given by
\begin{equation}
F_{\gamma/e}  (x,\xi) \equiv \frac{1}{\sigma_c} \frac{d\sigma_c}{dx} =
\frac{1}{D(\xi)} \left[ 1 - x + \frac{1}{1-x} - \frac{4x}{\xi (1-x)} +
\frac{4
x^2}{\xi^2 (1-x)^2}  \right] \; ,
\label{f:l}
\end{equation}
with
\begin{equation}
D(\xi) = \left(1 - \frac{4}{\xi} - \frac{8}{\xi^2}  \right) \ln (1 + \xi) +
\frac{1}{2} + \frac{8}{\xi} - \frac{1}{2(1 + \xi)^2} \; ,
\end{equation}
where $\sigma_c$ is the Compton cross section, $\xi \simeq 4
E\omega_0/m_e^2$,  $m_e$ and $E$ are the electron mass and
energy respectively, and $\omega_0$ is the laser-photon
energy. The fraction $x$ represents the ratio between the
scattered photon and initial electron energy for  the
backscattered photons traveling along the initial electron
direction. The maximum value of $x$ is
\begin{equation}
x_{\text{max}} = \frac{\omega_{\text{max}}}{E}
= \frac{\xi}{1+\xi} \; ,
\end{equation}
with $\omega_{\text{max}}$ being the maximum scattered
photon energy.

The fraction of photons with energy close to the maximum value grows
with $\sqrt{s}$ and $\omega_0$. Nevertheless, the bound $\xi < 2(1 +
\sqrt{2})$ should be respected in order to avoid the reduction in the
efficiency of the $e\to\gamma$ conversion due to the creation of
$e^+e^-$ pairs in collisions of the laser with backscattered photons.
We assumed that $\omega_0$ has the maximum value compatible with the
above constraint, {\em e.g.\/} for $\sqrt{s} = 500 $ GeV, $\omega_0 =
1.26$ eV and $x_{\text{max}} \simeq 0.83$. With this choice, more than
half of the scattered photons are emitted inside a small angle
($\theta < 5 \times 10^{-6}$ rad) and carry a large amount of the
electron energy. Due to this hard photon spectrum, the luminosity Eq.\
(\ref{lum}) is almost constant for $z < x_{\text{max}}$.

The analytical calculation of the cross section for the process
$\gamma\gamma \rightarrow W^+W^- \gamma$ ($\gamma\gamma \rightarrow
W^+W^-Z^0$) requires the evaluation of twelve Feynman diagrams in the
unitary gauge, which is a tedious and lengthy calculation despite of
being straightforward.  For the sake of completeness, we exhibit in
the Appendix the expression of the amplitudes of these processes.  In
order to perform these calculations in a efficient and reliable way
\cite{red}, we used an improved version of the numerical technique
presented in Ref.\ \cite{bar:num,zep:num}. The integrations were also
performed numerically using a Monte Carlo routine \cite{lepage} and we
tested the Lorentz and $U(1)_{\text em}$ gauge invariances of our
results for the amplitudes.

\section{Cross Sections and Gauge-boson Distributions}
\label{cs:dis}

We have evaluated the total cross section for the processes
$\gamma\gamma \rightarrow W^+W^-V$ imposing kinematical cuts on the
final state particles.  Our first cut  required that the produced
gauge bosons are in the central region of the detector, {\em i.e.\/} we
imposed that the angle of vector boson with the beam pipe is larger
than $30^\circ$, which corresponds to a cut in the pseudo-rapidity of
$|\eta| < 1.32$.  We further required the isolation of the final
particles by demanding that all vector bosons make an angle larger than
$25^\circ$ between themselves. Moreover, for the process \ref{g}, we
imposed a cut on the photon transverse momentum, $p_T^\gamma >10$ GeV,
to guarantee that the results are free of infrared divergences and to
mimic the performance of a typical electromagnetic calorimeter.

In Tables \ref{total:z} and \ref{total:g} we exhibit the results for
the total cross section of the processes \ref{z} and \ref{g}, with and
without the above cuts. As we can see from these tables, the
two-gauge-boson cross section ($\gamma + \gamma \rightarrow W^+ +
W^-$), which is the main reaction in a $\gamma\gamma$ collider
\cite{gin}, is from 2 to 4 orders of magnitude above those for three
gauge bosons depending upon $\sqrt{s}$. Nevertheless, we still find
promising event rates for final states $W^+W^-V$ for an $e^+e^-$
collider with an annual integrated luminosity of 10 fb$^{-1}$.
Moreover, the triple-gauge-boson production in $e^+e^-$ and
$\gamma\gamma$ colliders are comparable at $\sqrt{s}=500$ GeV, while
the event rate in $\gamma\gamma$ collider is a factor of 5--10 larger
than the one in a $e^+e^-$ machine at $\sqrt{s}=1$ TeV. The observed
growth of the total cross section for the production of three gauge
bosons is due to gauge-boson exchange in the $t$ and $u$ channels.

Since we are interested in final states where all the gauge  bosons are
identified, the event rate is determined not only by the total cross
section, but also by the  reconstruction efficiency that depends on the
particular decay channels of the vector bosons. In principle,  charged
lepton and light quark jet  pairs can be easily identified.
However, in the semileptonic decay of heavy quark the presence of
unmeasurable neutrinos spoils the invariant mass measurement, and we
adopt, as in Ref.\ \cite{bar:num}, that the efficiency for
reconstruction of a $W^\pm$ ($Z^0$) is 0.61 (0.65). In general,
final-state photons can be identified  with high efficiency as an
electromagnetic shower with a neutral initiator. Combining the
reconstruction efficiencies for individual particles, we obtain that the
process \ref{z} (\ref{g}) has a detection efficiency of 0.24 (0.37).
Once the reconstruction efficiency is substantial, the crucial factor for
event rates is the production cross section. Assuming the above cuts and
efficiencies we expect, for a 500 (1000) GeV collider with an annual
integrated luminosity of $10$ fb$^{-1}$, a total yield of 25 (198)
$\gamma + \gamma \rightarrow W^+ + W^- + Z^0$  fully reconstructed
events per year and 428 (714) $\gamma + \gamma \rightarrow W^+ + W^- +
\gamma$ reconstructed events per year with $P^\gamma_T > 10$ GeV.

In order to reach a better understanding of these reactions, we
present in Fig.\ \ref{fig:1}--\ref{fig:6} various distributions of the
final state gauge bosons.  In Fig.\ \ref{fig:1} we show the
distribution in $\cos\theta$, where $\theta$ is the polar angle of the
particles ($W^\pm$, and $V= \gamma, Z^0$) with the beam pipe. The
results are presented with and without the angular cuts described
above. The $W^+$ and $W^-$ curves coincide due to the charge
conjugation invariance.  We should notice that these processes are
particularly sensitive to central region requirement since,
analogously to what happens in the reaction $\gamma\gamma \rightarrow
W^+ W^-$, the $W$'s go preferentially along the beam pipe direction. This
fact can also be seen from the rapidity distribution of the final
state particles (Fig.\ \ref{fig:2}).  Therefore, the requirement that
the gauge bosons are produced in the central region of the detector
implies in a loss of $1/2$ to $5/6$ of the total number of events.
Increasing the center of mass energy, the $W$'s tend to populate the
high rapidity region while the $V= \gamma, Z^0$ distribution maintains
its shape. Consequently, the cut in the $W$ angle with beam pipe
discards most of the high energy events.

In order to estimate the importance of the isolation cut on the final
particles, we present in Fig.\ \ref{fig:3} the distributions in the
angle between the vector bosons. Charge conjugation invariance of the
processes implies that the distribution for $W^+Z^0$ and $W^-Z^0$ are
the same. In both processes \ref{z} and \ref{g}, the $W$'s tend to be
back-to-back, while the $WV$ ($V=Z^0$ or $\gamma$) is relatively flat,
demonstrating that the isolation cut is not very restrictive.  The
distribution for different energies of the collider are quite similar,
apart from a constant factor due to the growth of the total cross
section.

The invariant mass distributions of the $W^+W^-$ and $W^\pm
Z^0~(\gamma)$ pairs are presented in Fig.\ \ref{fig:4}.  Once again
the $W^+Z^0~(\gamma)$ and $W^-Z^0~(\gamma)$ curves coincide. From this
Figure we can learn that the average invariant mass of the pairs
$W^+W^-$ is higher then the one for $WZ^0~(\gamma)$ pairs. As the
center of mass energy of the collider is increased the distributions
grows due to the growth of the total cross section. Moreover, the
invariant mass distribution for $WZ^0~(\gamma)$ and $W^+W^-$ pairs are
considerably different: the former is rather narrow and peaked at
small invariant masses while the later one is broader and peaked at
high invariant masses.

Figure \ref{fig:5} shows the laboratory energy distributions the
of the $W^\pm$ and $Z^0~(\gamma)$ gauge bosons. In the process
$\gamma\gamma \rightarrow W^+W^-Z^0$, the $E_Z$ and $E_{W^\pm}$
distributions are rather similar, with the average energy of the
$W^\pm$ being larger than the average $Z^0$ energy. As the center
of mass energy of the collider is increased the distributions
grow and become rather isolated, while the peaks broaden
systematically. In the process $\gamma\gamma \rightarrow W^+W^-
\gamma$, the distributions in $E_\gamma$ and $E_{W^\pm}$ are
very different due to the infrared  divergences: the $E_\gamma$
is strongly peaked towards small energies while $E_{W^\pm}$ is
rather broad and peaked at high energies. With the increase of
the collider energy the difference between the distribution
become clearer.

We exhibit in Fig.\ \ref{fig:6} the transverse-momentum distribution
for the $W^\pm$ and $Z^0~(\gamma)$ vector bosons.  There are no
distinctive difference between the distribution for $W^\pm$ and $Z^0$
in process \ref{z}, apart from the fact that the $Z^0$'s exhibit a
smaller average $p_T$ than the $W$'s. In the case of process \ref{g},
the distributions for $\gamma$ and $W^\pm$ are very different since
the first is peaked at very small $p_T$ due to the infrared
divergences.

{\it Note added.~} After completing this work, we came across an
estimate of the total elementary cross section for the processes
studied here done by M.\ Baillargeon and F.\ Boudjema \cite{boudjema}.

\acknowledgments
This work was partially supported by Conselho Nacional de
Desenvolvimento Cient\'{\i}fico e Tecnol\'ogico (CNPq), and
by  Funda\c{c}\~ao de Amparo \`a Pesquisa do Estado de S\~ao
Paulo (FAPESP).


\appendix

\section*{}

We collect in this appendix the expressions for the amplitudes of the
processes $\gamma\gamma \rightarrow W^+W^-V$, with $V=Z^0$ or $\gamma$.
The Feynman diagrams contributing to these processes are given in Fig.\
\ref{fig:7}. The momenta and polarizations of the initial photons where
denoted by ($k_1$, $k_2$) and ($\epsilon_\mu(k_1)$,
$\epsilon_\nu(k_2)$), while the momenta and polarizations of the final
state $W^+$, $W^-$ and $V$ are given by ($p_+$, $p_-$, $k_3$) and
($\epsilon_\alpha(p_+)$, $\epsilon_\beta(p_-)$, $\epsilon_\gamma(k_3)$)
respectively. For a given choice of the initial and final polarizations
the amplitude of these processes can be written as
\begin {equation}
M={G_{v} \epsilon_{\mu}(k_1)\epsilon_{\nu}(k_2)\epsilon_{\alpha}
(p_+)
\epsilon_{\beta}(p_-)\epsilon_{\gamma}(k_3)M_T^{\mu\nu\alpha\beta\gamma}}\;,
\end {equation}
with
\begin {equation}
M_T^{\mu\nu\alpha\beta\gamma}=\sum_{\imath=1}^{7}
{M_{\imath}^{\mu\nu\alpha\beta\gamma}}
\; ,
\end {equation}
where the ${M_{\imath}^{\mu\nu\alpha\beta\gamma}}$ is the
contribution of the set of diagrams $\imath$ to the processes.
The factor $G_v$ depends upon the process, assuming the value
$e^3$   for the production of $W^+W^-\gamma$ and the value $e^3
\cot^2{\theta_W}$, with $\theta_W$ being the Weinberg angle, for
the final state  $W^+W^-Z^0$.

In order to write a compact expression for the amplitude, it is
convenient to define the triple-gauge-boson coupling coefficient as
\begin {equation}
\Gamma_3^{\alpha\beta\gamma}(P_1,P_2)=\left[(2P_1+P_2)^{\beta}g^{\alpha\gamma}
-(2P_2+P_1)^{\alpha}g^{\beta\gamma}+(P_2-P_1)^{\gamma}
g^{\beta\alpha}\right]\;,
\end {equation}
the quartic-gauge-boson coupling
\begin {equation}
\Gamma_4^{\mu\nu\alpha\beta}=g^{\mu\alpha}g^{\nu\beta}+
g^{\mu\beta} g^{\nu\alpha}-2g^{\mu\nu}g^{\alpha\beta}\; ,
\end {equation}
and the  propagator tensor
\begin {equation}
D^{\mu\nu}(k)=
\frac{(g^{\mu\nu} - k^\mu k^\nu / m^2)}{k^2-m^2} \;.
\end {equation}

Using the above definitions, the contributions of the different  set of
diagrams can be written as
\begin {eqnarray}
M_1^{\mu\nu\alpha\beta\gamma}&=&\Gamma_3^{\alpha\gamma\xi}(p_+,k_3)
D_{\xi\sigma}(p_++k_3)\Gamma_3^{\mu\sigma\rho}(k_1,-(p_++k_3))\nonumber
\\ & &
D_{\rho\lambda}(p_--k_2)\Gamma_3^{\beta\nu\lambda}(-p_-,k_2)
+ \; [ k_{1\leftrightarrow 2}\; ; \;\mu\leftrightarrow\nu ]
\end{eqnarray}
\begin {eqnarray}
M_2^{\mu\nu\alpha\beta\gamma}&=&
\Gamma_3^{\alpha\beta\xi}(k_3,p_-)
D_{\xi\sigma}(p_-+k_3)\Gamma_3^{\sigma\nu\rho}(-p_--k_3,k_2)
\nonumber \\
& &
D_{\rho\lambda}(k_1-p_+)\Gamma_3^{\mu\alpha\lambda}(-p_+,k_2)
+\; [ k_{1\leftrightarrow 2}\; ; \;\mu\leftrightarrow\nu ]
\end {eqnarray}
\begin {eqnarray}
M_3^{\mu\nu\alpha\beta\gamma}&=&\Gamma_3^{\mu\alpha\xi}(k_1,-p_+)
D_{\xi\sigma}(k_1-p_+)\Gamma_3^{\gamma\sigma\rho}
(-k_3,(k_1-p_+))
\nonumber \\
& &
D_{\rho\lambda}(p_--k_2)\Gamma_3^{\nu\beta\lambda}(-k_2,p_-)
+\; [ k_{1\leftrightarrow 2}\; ; \;\mu\leftrightarrow\nu ]
\end {eqnarray}
\begin {eqnarray}
M_4^{\mu\nu\alpha\beta\gamma}=\Gamma_3^{\beta\nu\xi}(-p_-,k_2)
D_{\xi\lambda}(k_2-p_-) \Gamma_4^{\lambda\alpha\mu\gamma}
+\; [ k_{1\leftrightarrow 2}\; ; \;\mu\leftrightarrow\nu ]
\end {eqnarray}
\begin {eqnarray}
M_5^{\mu\nu\alpha\beta\gamma}=\Gamma_3^{\mu\alpha\xi}(k_1,-p_+)
D_{\xi\lambda}(k_1-p_+) \Gamma_4^{\lambda\beta\nu\gamma}
+\; [ k_{1\leftrightarrow 2}\; ; \;\mu\leftrightarrow\nu ]
\end {eqnarray}
\begin {eqnarray}
M_6^{\mu\nu\alpha\beta\gamma}=
\Gamma_3^{\alpha\gamma\xi}(p_+,k_3)
D_{\xi\lambda}(p_++k_3)\Gamma_4^{\lambda\beta\nu\mu}
\end {eqnarray}
\begin {eqnarray}
M_7^{\mu\nu\alpha\beta\gamma}=
\Gamma_3^{\gamma\beta\xi}(k_3,p_-)
D_{\xi\lambda}(-p_--k_3)\Gamma_4^{\lambda\alpha\nu\mu}
\end {eqnarray}
where $[ k_{1\leftrightarrow 2}\; ; \;\mu\leftrightarrow\nu ]$
indicates the crossed contributions of the initial photons.



\begin{table}
\caption{Total cross section in fb for the process $\gamma \gamma
\rightarrow W^+ W^-Z^0$.}
\label{total:z}
\begin{tabular}{||c|c|c||}
$\sqrt{s}$ (GeV) & without cuts & with cuts \\
\tableline
500  &  20.4 & 10.2 \\
1000 & 289 & 81.9 \\
\end{tabular}
\end{table}

\begin{table}
\caption{Total cross section in fb for the process $\gamma \gamma
\rightarrow W^+W^-\gamma$}
\label{total:g}
\begin{tabular}{||c|c|c|c|c||}
&\multicolumn{2}{c|}{$P_T^\gamma > 10$ GeV}
&\multicolumn{2}{c||}{$P_T^\gamma > 20$ GeV}  \\
\hline
$\sqrt{s}$ (GeV) & without cuts & with cuts & without cuts & with cuts \\
\tableline
500  &  296 & 115 & 167 &  69 \\
1000 & 1162 & 192 & 748 & 138 \\
\end{tabular}
\end{table}


\protect
\begin{figure}
\protect
\caption{Angular distributions of the vector bosons with the beam pipe.
The upper (lower) solid lines stand for the W's, while the upper (lower)
dashed line represents the $V$ ($V=Z^0$ or $\gamma$) without (with) the
cuts discussed in the text. For the $W^+W^-\gamma$ production we imposed
the cut $p_T^\gamma >10$ GeV.}
\label{fig:1}
\end{figure}

\begin{figure}
\protect
\caption{Rapidity distributions. The conventions are the
same as Fig.\ \protect\ref{fig:1}.
}
\label{fig:2}
\end{figure}

\begin{figure}
\protect
\caption{Distributions of the angles between the pair of vector bosons.
The upper (lower) solid line stands for the $W^+W^-$ angle while the
upper (lower) dashed line represents the $WV$ angle without (with) the
cuts discussed in the text.  For the $W^+W^-\gamma$ production we
imposed the cut $p_T^\gamma >10$ GeV.}
\label{fig:3}
\end{figure}

\begin{figure}
\protect
\caption{Invariant mass distributions. The conventions are the
same as Fig.\ \protect\ref{fig:3}.}
\label{fig:4}
\end{figure}

\begin{figure}
\protect
\caption{Energy distributions. The conventions are the
same as Fig.\ \protect\ref{fig:1}.}
\label{fig:5}
\end{figure}

\begin{figure}
\protect
\caption{Transverse momentum distributions. The conventions are the
same as Fig.\ \protect\ref{fig:1}.}
\label{fig:6}
\end{figure}

\begin{figure}
\protect
\caption{Feynman diagrams that contribute to the process
$\gamma\gamma \rightarrow W^+W^-V$ with $V=Z^0$ or $\gamma$.}
\label{fig:7}
\end{figure}


\begin{references}

\bibitem{ano:ee} K.\ Hagiwara, R.\ D.\ Peccei, D.\ Zeppenfeld, and
K.\ Hikasa, Nucl.\ Phys.\ {\bf B282}, 253 (1987); Proceedings of the
ECFA Workshop on LEP200, CERN Report 87-08, ECFA Report 87/108, ed.\
A.\ B\"ohm and W.\ Hoogland, Aachen 1987.

\bibitem{bar:plb} V.\ Barger and T.\ Han, Phys.\ Lett.\ {\bf 212B}, 117
(1988).

\bibitem{bar:num} V.\ Barger, T.\ Han, and R.\ J.\ N.\ Phillips, Phys.\
Rev.\ D {\bf 39}, 146 (1989).

\bibitem{gunion} A.\ Tofighi-Niaki and J.\ F.\ Gunion, Phys.\
Rev.\ D {\bf 39}, 720 (1989).

\bibitem{golden} M.\  Golden and S.\ Sharpe, Nucl.\ Phys.\ {\bf B261},
217 (1985).

\bibitem{las0} F.\ R.\ Arutyunian, and V.\ A.\ Tumanian, Phys.\ Lett.\ {\bf
4}, 176 (1963);  R.\ H.\ Milburn, Phys.\ Rev.\ Lett.\ {\bf 10}, 75 (1963);
see also  C.\ Akerlof, University of Michigan Report No. UMHE 81-59 (1981),
unpublished.

\bibitem{laser} I.\ F.\ Ginzburg, G.\ L.\ Kotkin, V.\ G.\ Serbo, and V.\ I.\
Telnov, Nucl.\ Instrum.\  Methods {\bf 205}, 47 (1983); {\bf 219}, 5
(1984);  V.\ I.\ Telnov, Nucl.\ Instrum.\ Methods {\bf A294}, 72 (1990).

\bibitem{pal} R.\ B.\ Palmer, Annu.\  Rev.\ Nucl.\ Part.\ Sci.\ {\bf 40},
529 (1990).

\bibitem{bur} D.\ L.\ Burke, ``Linear Colliders: When? -- How?", to appear
in the Proceedings of the XXVI International Conference on High Energy
Physics", Dallas (1992).

\bibitem{strong} M.\ Chanowitz and M.K.\ Gaillard, Phys.\ Lett.\
{\bf 142B}, 85 (1984).

\bibitem{red} We have also checked our results using the analytical
amplitudes obtained using REDUCE and FORM. It turned out that the
numerical evaluation of cross sections and distributions by the
numerical technique is at least ten times faster than the analytical
results obtained with REDUCE and FORM.

\bibitem{zep:num}  K.\ Hagiwara, and D.\ Zeppenfeld, Nucl.\ Phys.\  {\bf B274},
1 (1986).

\bibitem{lepage} G.\ P.\ Lepage, J.\ Comp.\ Phys.\ {\bf 27}, 192 (1978).

\bibitem{gin} I.\ F.\  Ginzburg, G.\ L.\ Kotkin, S.\ L.\ Panfil, and
V.\ G.\ Serbo, Nucl.\ Phys.\ {\bf B228}, 285 (1983).

\bibitem{boudjema}  M.\ Baillargeon and F.\ Boudjema in Proceedings
of the ``Beyond the Standard Model III", Ottawa, Ontario, June 1992,
edited by S.\ Godfrey (World Scientific).

\end{references}
\end{document}